\documentclass[aps,preprint,nofootinbib]{revtex4}%
\usepackage{amsfonts}
\usepackage{amsmath}
\usepackage{amssymb}
\usepackage{graphicx}%
\providecommand{\U}[1]{\protect\rule{.1in}{.1in}}

\begin{document}
\preprint{ }
\title[Short title for running header]{Remarks on \textquotedblleft Note about Hamiltonian formalism of healthy
extended Ho\v{r}ava-Lifshitz gravity\textquotedblright\ \ by J. Kluso\v{n} }
\author{N. Kiriushcheva}
\email{nkiriush@uwo.ca}
\author{P. G. Komorowski}
\email{pkomoro@uwo.ca}
\author{S. V. Kuzmin}
\email{skuzmin@uwo.ca}
\affiliation{The Department of Applied Mathematics, The University of Western Ontario,
London, Ontario, N6A 5B7, Canada}

\begin{abstract}
We reassess the conclusion by Kluso\v{n} (\textit{J. High Energy Phys. 1007
(2010) 038}) that the Hamiltonian formulation of the healthy extended
Ho\v{r}ava-Lifshitz gravity does not present any problem.

\end{abstract}
\volumeyear{year}
\volumenumber{number}
\issuenumber{number}
\eid{identifier}
\maketitle

\section{Introduction}

In note \cite{KlusonJHEP} and its companion note \cite{KlusonPRD} the author
applies the Hamiltonian formalism to the actions of the so-called healthy
extension of the Ho\v{r}ava and Ho\v{r}ava-type models \cite{HoravaJHEP2009}.
The healthy extension was proposed in \cite{BlasPulojasSibiryakovPRL} where,
in particular, it was claimed that the canonical structure (Hamiltonian
formulation) of Ho\v{r}ava-type actions supplemented by the healthy extension
does not present any problem, contrary to the occurrence of pathologies in the
Hamiltonian formulation of Ho\v{r}ava's original proposal for the projectable
and non-projectable cases (e.g. see \cite{MP, HKG1, Pons, HKG}). Both papers
\cite{KlusonJHEP, KlusonPRD} confirm the assertion of
\cite{BlasPulojasSibiryakovPRL} about the health of the healthy extension in
the Hamiltonian formalism; we shall reassess this conclusion.

\section{Re-examination of Kluso\v{n}'s Hamiltonian analysis}

\subsection{Brief review of Kluso\v{n}'s notation and results}

The Hamiltonian formulation of the healthy extended Ho\v{r}ava models was
considered in \cite{KlusonJHEP} and for a slightly more complicated model in
\cite{KlusonPRD}; the latter produces nothing peculiar in comparison with the
simpler model. The action given by%
\begin{equation}
S\left(  N,N^{i},g_{km}\right)  =\int dtd^{D}x\sqrt{g}N\left(  K_{ij}%
G^{ijkl}K_{kl}-E^{ij}G_{ijkl}E^{kl}-V\left(  g_{ij},a_{i}\right)  \right)
\text{ }\label{eqnK5}%
\end{equation}
(see Eq. (2.3) of \cite{KlusonJHEP}), where%
\begin{equation}
K_{ij}=\frac{1}{2N}\left(  \dot{g}_{ij}-\nabla_{i}N_{j}-\nabla_{j}%
N_{i}\right)  ,\label{eqnK6}%
\end{equation}%
\begin{equation}
E^{ij}=\frac{\delta W}{\delta g_{ij}},\label{eqnK7}%
\end{equation}
and $G_{ijkl}$ is inverse of%
\begin{equation}
G^{ijkl}=\frac{1}{2}\left(  g^{ik}g^{jl}+g^{il}g^{jk}\right)  -\lambda
g^{ij}g^{kl}\label{eqnK8}%
\end{equation}
for $\lambda\neq\frac{1}{D}$, where $D$ is spatial dimension (more details can
be found in any article on Ho\v{r}ava-type models). The covariant derivative
with respect to the spatial metric is denoted by $\nabla_{j}$. The potential
$V\left(  g_{ij},a_{i}\right)  $ was not specified in either \cite{KlusonJHEP}
or \cite{KlusonPRD}, but the author refers to the paper in which it was
introduced (see Eq. (6) of \cite{BlasPulojasSibiryakovPRL}), where the
particular choice of terms was also explained. We reproduce the equation for
$V\left(  g_{ij},a_{i}\right)  $ here:
\begin{equation}
V\left(  g_{ij},a_{i}\right)  =-\alpha g^{ij}a_{i}a_{j}+M_{P}^{-2}\left(
C_{1}a_{i}\bigtriangleup a^{i}+C_{2}\left(  a_{i}a^{i}\right)  ^{2}+C_{3}%
a_{i}a_{j}R^{ij}+...\right)  +\label{eqnK10}%
\end{equation}%
\[
M_{P}^{-4}\left(  D_{1}a_{i}\bigtriangleup^{2}a^{i}+D_{2}\left(  a_{i}%
a^{i}\right)  ^{3}+D_{3}a_{i}a^{i}a_{j}a_{k}R^{jk}+...\right)
\]
with%
\begin{equation}
\bigtriangleup=g^{ij}\nabla_{i}\nabla_{j}\label{eqnK11}%
\end{equation}
and%
\begin{equation}
a_{i}\equiv\frac{\partial_{i}N}{N}=\partial_{i}\ln N.\label{eqnK12}%
\end{equation}
Note: although $a_{i}$ is sometimes called an \textquotedblleft additional
vector\textquotedblright\cite{KlusonPRD}, it is only a short notation for
combination (\ref{eqnK12}), not an independent variable. This combination is a
\textquotedblleft$D$-dimensional vector\textquotedblright\cite{KlusonJHEP} or
\textquotedblleft3-vector\textquotedblright\ in the original paper
\cite{BlasPulojasSibiryakovPRL}.

The first steps of the Hamiltonian formulation are standard and were followed
in \cite{KlusonJHEP}. By performing the Legendre transformation,
\[
H=\dot{N}p_{N}+\dot{N}^{i}p_{i}+\dot{g}_{ij}p^{ij}-L,
\]
where $p_{N}$, $p_{i}$, $p^{ij}$ are the momenta conjugate to all independent
variables $N$, $N^{i}$, $g_{ij}$ of (\ref{eqnK5}), and expressing the
velocities in terms of momenta, one obtains the following total
Hamiltonian\footnote{A non-standard definition is used in \cite{KlusonJHEP} --
\textquotedblleft the total Hamiltonian is the sum of the original Hamiltonian
and all constraints\textquotedblright; in the literature on constrained
dynamics such a combination is called an extended Hamiltonian, but this
extension is not healthy because equivalence with the Lagrangian is lost (see
\cite{HTbook}). For GR, the total Hamiltonian is a linear combination of
constraints, which is a common feature of generally covariant theories.}:%
\begin{equation}
H_{T}=\int d^{D}x\left[  N\left(  \mathcal{H}_{\perp}+\sqrt{g}V\right)
+N^{i}\mathcal{H}_{i}+\dot{N}^{i}p_{i}+\dot{N}p_{N}\right]  ,\label{eqnK15}%
\end{equation}
where $p_{i}$ and $p_{N}$ are primary constraints, and%
\begin{equation}
\mathcal{H}_{\perp}=\frac{1}{\sqrt{g}}p^{ij}G_{ijkl}p^{kl}+\sqrt{g}%
E^{ij}G_{ijkl}E^{kl},\label{eqnK16}%
\end{equation}%
\begin{equation}
\mathcal{H}_{i}=-2g_{ik}\partial_{j}p^{kj}-\left(  2\partial_{j}%
g_{ik}-\partial_{i}g_{jk}\right)  p^{jk}.\label{eqnK17}%
\end{equation}
Note that $\mathcal{H}_{\perp}$ is the same as the so-called Hamiltonian
constraint\footnote{In \cite{KlusonJHEP} a different notation is used for our
$\mathcal{H}_{\perp}$ (i.e. $\mathcal{H}_{T}$), which is somewhat confusing
since the standard use of subscript, $_{T}$, is for the total Hamiltonian.} of
the unhealthy Ho\v{r}ava model, and $\mathcal{H}_{i}$ coincides with the
momentum constraint for GR in ADM variables. The time development of the
primary constraints leads to%
\begin{equation}
\dot{p}_{i}=\left\{  p_{i},H_{T}\right\}  =-\mathcal{H}_{i}\approx
0,\label{eqnK20}%
\end{equation}%
\begin{equation}
\dot{p}_{N}=\left\{  p_{N},H_{T}\right\}  =-\mathcal{H}_{\perp}-\sqrt
{g}V+\frac{1}{N}\partial_{i}\left(  N\sqrt{g}\frac{\delta V}{\delta a_{i}%
}\right)  \equiv\Theta_{2}\approx0.\label{eqnK21}%
\end{equation}

Further, the consideration of the time development of (\ref{eqnK21}),%
\[
\dot{\Theta}_{2}=\left\{  \Theta_{2},H_{T}\right\}  =\left\{  \Theta
_{2},N\left(  \mathcal{H}_{\perp}+\sqrt{g}V\right)  +N^{i}\mathcal{H}%
_{i}\right\}  +\left\{  -\sqrt{g}V+\frac{1}{N}\partial_{i}\left(  N\sqrt
{g}\frac{\delta V}{\delta a_{i}}\right)  ,\dot{N}p_{N}\right\}  ,
\]
gives a non-linear partial differential equation for undetermined velocity,
$\dot{N}$ (the Lagrange multiplier). The primary, $p_{N}$, and secondary,
$\Theta_{2}$, constraints were classified in \cite{KlusonJHEP} as a
second-class pair, and this fact was presented as a key advantage of the
healthy extension because it allows one to perform the Hamiltonian reduction,
i.e. go to the reduced phase-space by eliminating the pair of canonical
variables ($N$, $p_{N}$): $p_{N}=0$ and $N=N\left(  \mathcal{H}_{\perp}%
,g_{km}\right)  $ (the solution to $\Theta_{2}=0$ for $N$ as a function of
remaining phase-space variables).

We note that solving $\Theta_{2}=0$ of (\ref{eqnK21}) is not the same as
finding a solution to the second-class constraints in known field-theoretical
examples, e.g. first-order formulation of the Maxwell, Yang-Mills theory, and
affine-metric formulation of GR, where such equations are algebraic with
respect to the eliminated fields, and the reduction can be performed either at
the Lagrangian or Hamiltonian levels, with the same outcome (e.g. see
\cite{AnnPhys}). This important difference was not commented on or even
noticed in \cite{KlusonJHEP, KlusonPRD}, and an assumption was made that the
solution to (\ref{eqnK21}) can be found. The appearance of differential
second-class constraints\footnote{In field theories, differential constraints
are not something unusual, but here \textquotedblleft
differential\textquotedblright\ manifests not just a presence of some
derivatives, but derivatives of fields with respect to which these constraints
should be solved.} is a common feature of Hamiltonian formulations of
Ho\v{r}ava-type models, but to the best of our knowledge, this fact was
mentioned only in a paper by Pons and Talavera \cite{Pons} (see footnote 3),
without elaboration on the effect or possible consequences of solving such
second-class constraints in the Hamiltonian reduction; it is merely mentioned,
that unlike mechanics, where the multipliers are \textquotedblleft undoubtedly
determined\textquotedblright, this is \textquotedblleft not the case in field
theory\textquotedblright. Therefore, for differential constraints like
(\ref{eqnK21}), a multiplier cannot be uniquely determined; the term
\textquotedblleft the partial determination of the
multiplier\textquotedblright\ is used in \cite{Pons}. If a multiplier is
partially determined, the constraint is only partially second-class;
therefore, it is also partially first-class, and the Hamiltonian reduction can
(if at all) only be partially performed. Problems of this sort, as well as the
problems in calculating the Dirac Brackets for such partial second-class
constraints, are discussed in \cite{Kurt} (see Section 4 of Chapter 2); but an
undoubted solution to these problems is unknown, and should be found before
any conclusion can be drawn about formulations that have differential
(partially) second-class constraints. If neglected, these problems will
reappear at the later stages of the Hamiltonian procedure. To demonstrate this
fact we shall continue our analysis, and like the author of \cite{KlusonJHEP,
KlusonPRD}, assume that such constraints have no peculiarities, and we return
to the equation:
\begin{equation}
-\mathcal{H}_{\perp}-\sqrt{g}V+\frac{1}{N}\partial_{i}\left(  N\sqrt{g}%
\frac{\delta V}{\delta a_{i}}\right)  =0.\label{eqnK25}%
\end{equation}
Note: equation (\ref{eqnK25}) is a complicated, nonlinear partial differential
equation of higher order (e.g. considering the term with $\bigtriangleup^{2}$
in (\ref{eqnK10}) makes this equation of sixth order in spatial derivatives of
lapse); and the existence and uniqueness of a solution deserves careful
analysis. As a second-class constraint, (\ref{eqnK25}) is assumed to have a
solution that allows one to perform the Hamiltonian reduction
\cite{KlusonJHEP}, i.e. one can find $N=N\left(  \mathcal{H}_{\perp}%
,g_{km}\right)  $ (express one canonical variable in terms of others). After
such an elimination it is also necessary to find the Dirac Brackets (again,
not a unique operation for differential constraints (see \cite{Kurt})); but if
one neglects these subtle questions and just mimics the case of mechanical
models, or algebraic constraints in field theory, then the result follows
immediately -- the Dirac Brackets, for the variables remaining after
reduction, are the same as the Poisson Brackets (PBs).

The solution of the second-class constraints (\ref{eqnK15}), i.e. $p_{N}=0$
and $N=N\left(  \mathcal{H}_{\perp},g_{km}\right)  $, must be substituted into
Hamiltonian; and the reduced total Hamiltonian follows,
\begin{equation}
H_{T}=\int d^{D}x\left[  N\left(  \mathcal{H}_{\perp},g\right)  \left(
\mathcal{H}_{\perp}+\sqrt{g}V\left(  \mathcal{H}_{\perp},g\right)  \right)
+N^{i}\mathcal{H}_{i}+\dot{N}^{i}p_{i}\right]  \label{eqnK26}%
\end{equation}
(see Eq. (2.33) of \cite{KlusonJHEP}).

Equation (\ref{eqnK26}) is the culmination, or end point, of the Hamiltonian
formulation in \cite{KlusonJHEP}. In the discussion after this equation the
author emphasizes that \textquotedblleft the Hamiltonian constraint is missing
in the healthy extended [case]\textquotedblright\ and the total Hamiltonian is
not given as a linear combination of constraints. But upon elimination of
second-class constraints, the Hamiltonian analysis is not complete; to find
the number of Degrees of Freedom (DoF) (although it can simplify the search
for closure, the elimination of the second-class constraints is not always
needed), and to restore gauge invariance using the Dirac conjecture (all
first-class constraints generate a gauge symmetry), one must demonstrate the
closure of the Dirac procedure for the reduced Hamiltonian.

\subsection{Continuation of Hamiltonian analysis}

The reduced total Hamiltonian (\ref{eqnK26}) was obtained under the assumption
that the constraints, $p_{N}=0$ and $\Theta_{2}=0$, are second class and can
be eliminated (i.e. a solution, $N=N\left(  \mathcal{H}_{\perp},g_{km}\right)
$ can be found). The next step is to check closure of the Dirac procedure in
the reduced phase-space, and consider the time development of the secondary
constraint (i.e. the PB of $\mathcal{H}_{i}$ with the reduced total
Hamiltonian),%
\[
\mathcal{\dot{H}}_{i}=\left\{  \mathcal{H}_{i}\left(  x\right)  ,\int
d^{D}yH_{T}\left(  y\right)  \right\}  =
\]%
\begin{equation}
\left\{  \mathcal{H}_{i},\int d^{D}yN\left(  \mathcal{H}_{\perp},g\right)
\left(  \mathcal{H}_{\perp}+\sqrt{g}V\left(  \mathcal{H}_{\perp},g\right)
\right)  \right\}  +\left\{  \mathcal{H}_{i},\int d^{D}yN^{k}\mathcal{H}%
_{k}\right\}  . \label{eqnK28}%
\end{equation}
The second PB of (\ref{eqnK28}) is known, and it is proportional to the
secondary constraints%
\begin{equation}
\left\{  H_{i},\int d^{D}yN^{m}H_{m}\right\}  =\partial_{i}N^{k}H_{k}%
+\partial_{k}\left(  N^{k}H_{i}\right)  . \label{eqnK29}%
\end{equation}
But the first PB of (\ref{eqnK28}) is more complicated, and only partial
results can be obtained easily, e.g. for the first contribution,
\[
\left\{  \mathcal{H}_{i}\left(  x\right)  ,\int d^{D}yN\left(  \mathcal{H}%
_{\perp},g\right)  \mathcal{H}_{\perp}\right\}  =\left\{  \mathcal{H}%
_{i}\left(  x\right)  ,\int d^{D}y[\mathcal{H}_{\perp}\left(  y\right)
\right\}  N\left(  \mathcal{H}_{\perp},g\right)  \left(  y\right)  ]+
\]%
\begin{equation}
\left\{  \mathcal{H}_{i}\left(  x\right)  ,\int d^{D}y[N\left(  \mathcal{H}%
_{\perp},g\right)  \left(  y\right)  \right\}  \mathcal{H}_{\perp}\left(
y\right)  ]=\partial_{i}N\left(  \mathcal{H}_{\perp},g\right)  \mathcal{H}%
_{\perp}+...~, \label{eqnK30}%
\end{equation}
where we used the known result for $\mathcal{H}_{\perp}$ of (\ref{eqnK16}),%
\[
\left\{  \mathcal{H}_{i}\left(  x\right)  ,\int d^{D}yf\left(  y\right)
\mathcal{H}_{\perp}\left(  y\right)  \right\}  =\partial_{i}f\left(  x\right)
\mathcal{H}_{\perp}\left(  x\right)  .
\]

To have spatial diffeomorphism gauge invariance, which is claimed to be the
gauge symmetry of the healthy extended action, the algebra of constraints must
be of a very special form and be closed on the secondary constraints. This
imposes severe restrictions on the first PB in (\ref{eqnK28}). If first PB is
not zero, then a tertiary constraint arises; if it is proportional to a
momentum constraint (not seems to be the case in (\ref{eqnK30})), then a
generator\footnote{The generator of spatial diffeomorphism is not just a
momentum constraint as it is often presented in the literature. If this were
true, then the gauge transformations of lapse and shift functions would be
zero, which is not the case, and this is a well known fact from the Lagrangian
and Hamiltonian analyses of such actions (e.g. see the generator for GR in ADM
variables \cite{Castellani}). In general, according to Dirac \textquotedblleft
if we are to have any motion at all with a zero Hamiltonian, we must have at
least one primary first-class constraint\textquotedblright\ \cite{Diracbook}%
.\ This fact is very often neglected in the Hamiltonian analysis of the ADM
formulation of GR and other ADM-inspired models.} will be changed, and the
gauge transformations will be different. The only result compatible with
spatial diffeomorphism is that the first PB in (\ref{eqnK28}) is equal to
zero. Is this the case? The answer to this question is important, both for the
restoration of gauge symmetry, and for DoF counting. In spite of the
complexity of expression (\ref{eqnK28}), the answer is: yes (although it
raises a new question, to which we shall return). The first PB of
(\ref{eqnK28}) is zero; this is true irrespective of the form of solution
(\ref{eqnK25}), and true for all terms of the potential given by
(\ref{eqnK10}) (compare the first term in (\ref{eqnK26}) with (\ref{eqnK25})).
Taking $\mathcal{H}_{\perp}$ from (\ref{eqnK25}) gives%
\[
N\left(  \mathcal{H}_{\perp},g\right)  \left(  \mathcal{H}_{\perp}+\sqrt
{g}V\left(  \mathcal{H}_{\perp},g\right)  \right)  =\partial_{i}\left(
N\sqrt{g}\frac{\delta V}{\delta a_{i}}\right)  _{N=N\left(  \mathcal{H}%
_{\perp},g\right)  };
\]
therefore, the first PB of (\ref{eqnK28}) is indeed zero, and the total
reduced Hamiltonian is actually%
\begin{equation}
H_{T}=\int d^{D}x\left[  N^{i}\mathcal{H}_{i}+\dot{N}^{i}p_{i}\right]  .
\label{eqnK35}%
\end{equation}

Note: the author's assertion (in \cite{KlusonJHEP}) that the total Hamiltonian
of the healthy extended actions is not given as a linear combination of
constraints is incorrect. The assumption about the existence of a solution to
the second-class constraints leads to a very simple reduced Hamiltonian
(\ref{eqnK35}), which has closure on the secondary first-class constraints and
a simple constraint algebra (\ref{eqnK29}). One can easily calculate DoF (we
provide the result for configurational space) based on the number of
constraints and the number of variables in reduced phase-space. After the
elimination of lapse, the number of independent fields and the number of
first-class constraints, becomes $\frac{\left(  D+1\right)  \left(
D+2\right)  }{2}-1$ and $2D$, which yields%
\[
\#DoF=\#fields-\#FCC=\frac{D\left(  D-1\right)  }{2},
\]
according to \cite{HTZ}.

Let us find the gauge invariance from the first-class constraints. The number
of gauge parameters (and symmetries) is known -- it is the same as the number
of first-class primary constraints, $p_{i}$. Hamiltonian (\ref{eqnK35}) is
just a particular, simplified case of the result known for GR in ADM
variables, for which the generator of the gauge transformations was obtained
long ago by Castellani \cite{Castellani} (who developed a theorem based upon
the Dirac conjecture \cite{Diracbook}, and developed the procedure of
restoration of gauge symmetry, which has been successfully applied to many
Hamiltonian formulations of gauge theories). For details of application of
this method to GR in ADM variables, we refer the reader to the original paper
\cite{Castellani} (some additional details can be found in \cite{Myths}). The
generator of spatial diffeomorphism, $G_{S}$, for the total Hamiltonian
(\ref{eqnK35}) with algebra of constraints (\ref{eqnK29}) is%
\begin{equation}
G_{S}=\int d^{D}x\left(  \dot{\xi}^{i}p_{i}+\xi^{i}\left(  H_{i}+\partial
_{i}N^{j}p_{j}+\partial_{j}\left(  N^{j}p_{i}\right)  \right)  \right)
.\label{eqnK40}%
\end{equation}

The form of the generator is uniquely defined and reflects the whole algebra
of constraints (the generator involves all first-class constrains according to
the Dirac conjecture). The transformations are calculated using (keeping
convention of \cite{Castellani})%
\[
\delta_{S}field=\left\{  field,G_{S}\right\}  .
\]

We obtain the spatial diffeomorphism gauge transformation for two fields,
\begin{equation}
\delta_{S}N^{k}=-\xi^{j}\partial_{j}N^{k}+N^{j}\partial_{j}\xi^{k}-\dot{\xi
}^{k},\label{eqnK43}%
\end{equation}%
\begin{equation}
\delta_{S}g_{km}=-\partial_{m}\xi^{i}g_{ik}-\partial_{k}\xi^{i}g_{im}-\xi
^{i}\partial_{i}g_{km}.\label{eqnK44}%
\end{equation}
(field $N$ is not part of the reduced phase-space and generator). Note that
$N^{k}$ does not transform as a vector under spatial diffeomorphism
transformations because of the last, extra term in (\ref{eqnK43}).
Transformations (\ref{eqnK43})-(\ref{eqnK44}) are equal to the ones presented
in Ho\v{r}ava's original paper \cite{HoravaJHEP2009} (up to an overall sign,
and for the standard choice of shift, $N^{k}$, in ADM formulation). The choice
of shift is very important in Hamiltonian formulation because it affects the
constraint algebra \cite{KKK-6}. For the non-projectable case (i.e. where
lapse is a function, $N\left(  x,t\right)  $, that depends on space and time,
not on time alone, i.e. $N\left(  t\right)  $), which is subjected to the
healthy extension considered in \cite{KlusonJHEP}, the lapse should also
transform under spatial diffeomorphism (see \cite{HoravaJHEP2009}),%
\begin{equation}
\delta_{S}N=-\xi^{i}\partial_{i}N.\label{eqnK45}%
\end{equation}

In \cite{KlusonJHEP} a few different results can be found for the
transformations of $N$ and $a_{i}$. The transformation of lapse under spatial
diffeomorphism, according to the second line of Eq. (2.10), is zero, i.e.
$\delta_{S}N=0$ (there are no terms with the corresponding gauge parameters).
If this were the case, the transformations of $a_{i}$ would also be zero,
$\delta_{S}a_{i}=0$ (because of (\ref{eqnK12})); but they are given in Eq.
(2.11) as $\delta_{S}a_{i}=-a_{j}\partial_{i}\xi^{j}$, which is not the
transformation of a vector under spatial diffeomorphism that should follow
from (\ref{eqnK12}) and (\ref{eqnK45}). The correct transformation of vector
$a_{i}$ appears only later (see Eq. (2.20)),%
\begin{equation}
\delta_{S}a_{i}=-a_{j}\partial_{i}\xi^{j}-\xi^{j}\partial_{j}a_{i}%
~,\label{eqnK50}%
\end{equation}
and in such a case, lapse must transform according to (\ref{eqnK45}).

The Hamiltonian formulation provides an algorithmic way of finding the gauge
transformations of all fields (that do not need to be specified \textit{a
priori}). To find the gauge transformation of lapse, which is not part of
reduced phase-space, one must go back one step to return to the second-class
constraint (\ref{eqnK25}), and use the transformation of the metric in
(\ref{eqnK44}) (shift is not present in (\ref{eqnK25})).

\subsection{Restoration of gauge transformations of lapse from the
second-class constraint}

In gauge theories with second-class constraints, the restoration of gauge
symmetries for reduced variables proceeds as follows: the solution to the
second-class constraints is an expression for one variable in terms of the
other variables that have survived reduction in the Hamiltonian, and for which
the gauge transformations are known.

As a simple example, we refer to \cite{KlusonPRD} where one scalar field, $A$,
was eliminated by solving a second-class constraint $A=F\left(  B\right)  $,
and the transformations of field $B$ were found from the reduced Hamiltonian:
$\delta_{S}B=$ $-\xi^{i}\partial_{i}B$. This information is enough to find the
transformation of field $A$:%
\begin{equation}
\delta_{S}A=\delta_{S}F\left(  B\right)  =\frac{\delta F\left(  B\right)
}{\delta B}\delta_{S}B=-\frac{\delta F\left(  B\right)  }{\delta B}\xi
^{i}\partial_{i}B=-\xi^{i}\partial_{i}F\left(  B\right)  =-\xi^{i}\partial
_{i}A. \label{eqnK55}%
\end{equation}

For the case under consideration, the explicit solution of (\ref{eqnK25}) for
$N$ is unknown; but the gauge transformation of this equation, as well as that
for lapse, can be found in way similar to that shown in (\ref{eqnK55}) by
using equation (\ref{eqnK25}), which one may \cite{KlusonJHEP}
\textquotedblleft presume can be explicitly solved\textquotedblright. To
shorten our notation, we shall call the solution to (\ref{eqnK25}) $\tilde
{N}\equiv N\left(  \mathcal{H}_{T},g_{km}\right)  $, and find $\delta
_{S}\tilde{N}$ by using (\ref{eqnK25}) and the known transformations
(\ref{eqnK43})-(\ref{eqnK44}). To restore the gauge transformation, the
explicit form of potential $V\left(  g_{ij},a_{i}\right)  $ is needed. Let us
consider only one simple term from (\ref{eqnK10}),%
\begin{equation}
V\left(  g_{ij},a_{i}\right)  =\sqrt{g}Ng^{pq}a_{p}a_{q}~\label{eqnK60}%
\end{equation}
(for the rest of the contributions the result is the same). With this form of
potential, equation (\ref{eqnK25}) becomes%
\begin{equation}
-\mathcal{H}_{\perp}-\sqrt{g}g^{pq}\frac{1}{\tilde{N}^{2}}\partial_{p}%
\tilde{N}\partial_{q}\tilde{N}+2\frac{1}{\tilde{N}}\partial_{i}\left(
\sqrt{g}g^{iq}\partial_{q}\tilde{N}\right)  =0\label{eqnK62}%
\end{equation}
(where we have substituted the solution, $\tilde{N}$).

Performing variation $\delta_{S}$ of (\ref{eqnK62}) one obtains%
\begin{equation}
-\delta_{S}\mathcal{H}_{\perp}-\delta_{S}\left(  \sqrt{g}g^{pq}\right)
\frac{1}{\tilde{N}^{2}}\partial_{p}\tilde{N}\partial_{q}\tilde{N}+2\sqrt
{g}g^{pq}\frac{1}{\tilde{N}^{3}}\partial_{p}\tilde{N}\partial_{q}\tilde
{N}\left(  \delta_{S}\tilde{N}\right)  -2\sqrt{g}g^{pq}\frac{1}{\tilde{N}^{2}%
}\partial_{p}\tilde{N}\partial_{q}\left(  \delta_{S}\tilde{N}\right)
\label{eqnK63}%
\end{equation}%
\[
-2\frac{1}{\tilde{N}^{2}}\partial_{i}\left(  \sqrt{g}g^{iq}\partial_{q}%
\tilde{N}\right)  \left(  \delta_{S}\tilde{N}\right)  +2\frac{1}{\tilde{N}%
}\partial_{i}\delta_{S}\left(  \sqrt{g}g^{iq}\right)  \partial_{q}\tilde
{N}+2\frac{1}{\tilde{N}}\sqrt{g}g^{iq}\partial_{i}\partial_{q}\left(
\delta_{S}\tilde{N}\right)  =0.
\]
To solve this equation for $\delta_{S}\tilde{N}$, we substitute the known
transformation of $\mathcal{H}_{\perp}$ (as a scalar density):%
\begin{equation}
\delta_{S}\mathcal{H}_{\perp}=-\partial_{i}\left(  \xi^{i}\mathcal{H}_{\perp
}\right)  .\label{eqnK64}%
\end{equation}
And the combination, $\sqrt{g}g^{iq}$, which transforms as%
\begin{equation}
\delta_{S}\left(  \sqrt{g}g^{iq}\right)  =-\sqrt{g}g^{iq}\partial_{p}\xi
^{p}+\sqrt{g}\left(  g^{qm}\partial_{m}\xi^{i}+g^{ip}\partial_{p}\xi
^{q}\right)  -\partial_{p}\left(  \sqrt{g}g^{iq}\right)  \xi^{p}%
,\label{eqnK65}%
\end{equation}
is presented in a form that is explicitly linear in the gauge parameter and
its first-order spatial derivatives. Using (\ref{eqnK64}) and (\ref{eqnK62}),
the first term of (\ref{eqnK63}) can be written as%
\begin{equation}
-\delta_{S}\mathcal{H}_{\perp}=\partial_{k}\left(  \xi^{k}\left(  -\sqrt
{g}g^{pq}\frac{1}{\tilde{N}^{2}}\partial_{p}\tilde{N}\partial_{q}\tilde
{N}+2\frac{1}{\tilde{N}}\partial_{i}\left(  \sqrt{g}g^{iq}\partial_{q}%
\tilde{N}\right)  \right)  \right)  .\label{eqnK66}%
\end{equation}

Equation (\ref{eqnK63}), upon substitution of (\ref{eqnK65}) and
(\ref{eqnK66}), allows one to find $\delta_{S}\tilde{N}$. Note: this is not an
equation to find a gauge parameter, it should work for all values of the
field-independent vector function, $\xi^{p}$. We must find $\delta_{S}%
\tilde{N}$, which should be consistent with all orders of gauge parameter (it
is linear, but with different order of derivatives). It is clear that the
highest order derivative of the gauge parameter (for this part of potential)
is two (the second last term in (\ref{eqnK63})), and the highest order
derivative of $\delta_{S}\tilde{N}$ is also two (the last term in
(\ref{eqnK63})). Therefore, the\textit{ }transformation of $\delta_{S}%
\tilde{N}$ should be linear in the gauge parameter (without derivatives), i.e.%
\begin{equation}
\delta_{S}\tilde{N}=X_{m}\xi^{m}.\label{eqnK70}%
\end{equation}

To find $X_{m}$, we substitute (\ref{eqnK70}) into the last term of
(\ref{eqnK63}) and keep only contributions with second-order derivatives of
the parameter; and we do the same with the second last term, thus%
\[
+2\frac{1}{\tilde{N}}\left[  \sqrt{g}g^{ip}\partial_{i}\partial_{p}\xi
^{q}\right]  \partial_{q}\tilde{N}+2\frac{1}{\tilde{N}}X_{m}\sqrt{g}%
g^{iq}\partial_{i}\partial_{q}\xi^{m}=0,
\]
which yields%
\begin{equation}
X_{m}=-\partial_{m}\tilde{N}.\label{eqnK74}%
\end{equation}

Using (\ref{eqnK70}) and steps similar to those in (\ref{eqnK55}), we obtain
transformation (\ref{eqnK45}). Of course, the rest of terms in (\ref{eqnK63}),
those linear in the gauge parameter and linear in the first-order derivative
of the gauge parameter, must be checked for consistency with (\ref{eqnK70})
and (\ref{eqnK74}); but this is not difficult to confirm by a straightforward
calculation. We performed this calculation for the simplest part of potential
(\ref{eqnK60}), and it is not hard to repeat for the rest of terms in
(\ref{eqnK10}).

In the conclusion of \cite{KlusonJHEP} one can read the statement:
\textquotedblleft it would be also extremely useful to find explicit
dependence $N$ on $\mathcal{H}_{T}$ and $g$\textquotedblright. From the
transformation properties of a solution, it must be a scalar, and it follows
that the explicit dependence of $N$ becomes very restricted -- only
combinations of $\mathcal{H}_{\perp}$ and $g_{kn}$ that form a scalar are
possible. For example, $\tilde{N}=\frac{\mathcal{H}_{\perp}}{\sqrt{g}}$, or
any function of this combination, $F\left(  \frac{\mathcal{H}_{\perp}}%
{\sqrt{g}}\right)  $, are scalars, but not a solution of (\ref{eqnK25}). In
fact, it seems to us that there is no combination, which can be constructed
form $\mathcal{H}_{\perp}$ and $g_{kn}$, that is simultaneously a scalar and a
solution of (\ref{eqnK25}); but we were not able to prove this in general.

\section{Conclusion. Pathologies of Hamiltonian formulation of healthy
extension}

The assertion \cite{BlasPulojasSibiryakovPRL} that the canonical structure
(Hamiltonian formulation) of the Ho\v{r}ava-type actions, when supplemented by
the healthy extension, does not present any problem, and that analysis by
Kluso\v{n} \cite{KlusonJHEP, KlusonPRD} has confirmed this assertion, are
unjustifiably optimistic; the Hamiltonian formulation actually exhibits many pathologies.

The reduced Hamiltonian of the healthy extended Ho\v{r}ava model
(\ref{eqnK35}), under the assumption that the second-class constraint can be
solved, is too simple, and contrary to the statement of \cite{KlusonJHEP}, it
is linear in constraints. Moreover, after Hamiltonian reduction is performed,
one should be able to return to the original Lagrangian or its equivalent form
by performing the inverse Legendre transformation,%
\begin{equation}
L=\dot{N}^{i}p_{i}+\dot{g}_{ij}p^{ij}-H_{T}=\dot{g}_{ij}p^{ij}-N^{i}%
\mathcal{H}_{i}.\label{eqnK80}%
\end{equation}
In (\ref{eqnK80}) there is no contribution quadratic in momenta, and the
momenta cannot be expressed in terms of velocities, preventing one from
returning to the Lagrangian.

The order in which reduction is performed (i.e. to reduce at the Lagrangian
level then go to the Hamiltonian, or go from the Lagrangian to the Hamiltonian
and then reduce at the Hamiltonian level) is interchangeable \cite{AnnPhys}.
Given the action of the healthy extension (\ref{eqnK5})%
\begin{equation}
S\left(  N,N^{i},g_{km}\right)  =\int dtd^{D}x\sqrt{g}\left(  \frac{1}%
{N}\tilde{K}_{ij}G^{ijkl}\tilde{K}_{kl}-NE^{ij}G_{ijkl}E^{kl}-NV\left(
g_{ij},a_{i}\right)  \right)  , \label{eqnK81}%
\end{equation}
(to have an explicit dependence on $N$, we have introduced $\tilde{K}%
_{ij}\equiv NK_{ij}$), one may perform a variation of $S\left(  N,N^{i}%
,g_{km}\right)  $ with respect to $N$ to obtain:%
\begin{equation}
\frac{\delta S}{\delta N}=-\sqrt{g}\frac{1}{N^{2}}\tilde{K}_{ij}G^{ijkl}%
\tilde{K}_{kl}-\sqrt{g}E^{ij}G_{ijkl}E^{kl}-\sqrt{g}V+\frac{1}{N}\partial
_{i}\left(  N\sqrt{g}\frac{\delta V}{\delta a_{i}}\right)  =0, \label{eqnK82}%
\end{equation}
which produces a result similar to (\ref{eqnK25}). Therefore, if the solution
of the second-class constraint is assumed to exist, then (\ref{eqnK82}) can
also be solved. Performing such a Lagrangian reduction (i.e. substituting
solution of (\ref{eqnK82}) $N=N\left(  \tilde{K}_{ij}G^{ijkl}\tilde{K}%
_{kl},E^{ij}G_{ijkl}E^{kl},g_{km}\right)  =\tilde{N}$ (similar to $N=N\left(
\mathcal{H}_{\perp},g_{km}\right)  $)) into (\ref{eqnK81}), yields%
\begin{equation}
S\left(  N^{i},g_{km}\right)  =\int dtd^{D}x\sqrt{g}\frac{2}{\tilde{N}}%
\tilde{K}_{ij}G^{ijkl}\tilde{K}_{kl}~,\text{ \ \ } \label{eqnK85}%
\end{equation}
or in an equivalent form,%
\begin{equation}
S\left(  N^{i},g_{km}\right)  =\text{\ }S=\int dtd^{D}x\left(  -2\sqrt
{g}\tilde{N}E^{ij}G_{ijkl}E^{kl}-2\sqrt{g}\tilde{N}V\left(  g_{ij}%
,a_{i}\right)  \right)  _{a_{i}=\partial_{i}\ln\tilde{N}}~. \label{eqnK86}%
\end{equation}

The Hamiltonian formulation of (\ref{eqnK85}) or (\ref{eqnK86}) should lead to
reduced Hamiltonian (\ref{eqnK35}). After performing the Legendre
transformation%
\[
H_{T}=\dot{N}^{i}p_{i}+\dot{g}_{ij}p^{ij}-L\left(  N^{i},g_{km}\right)
\]
and then eliminating the velocities in terms of momenta. To obtain
(\ref{eqnK35}), the contributions, which are quadratic in momenta, must
disappear from a Lagrangian that is quadratic in velocities; but this is impossible.

In the Hamiltonian formulations of extended Ho\v{r}ava models
\cite{KlusonJHEP}, all problems originate from the assumption that the
equation for the differential (second-class) constraint can be solved, as is
the case for second-class algebraic constrains in the first-order formulations
of field theories (e.g. Maxwell, Yang-Mills, and the affine-metric formulation
of GR); but it is this appearance of such a differential constraint that has
been presented as an advantage of the extended formulation. The assumption
that a solution of such constraints exists is not specific to extended models,
and such differential constraints can be found in all Hamiltonian formulations
of Ho\v{r}ava-type models. This important difference and the complications
related to it \cite{Kurt} were never mentioned in the literature of the
Hamiltonian formulation of Ho\v{r}ava-type actions, with the exception of
\cite{Pons}; although it is good that this difference was recognized, one must
perform further analysis before any conclusion can be drawn\ about the
healthiness of the Hamiltonian formulations for models where such constraints
arose. Is this a problem that is related to some subtle points of the
Hamiltonian formulation of field-theoretical models (e.g. see
\cite{SeilerTucker}) or is it an indication of some innate pathologies in the
models? This is a question that must be answered.

The healthy extension \cite{BlasPulojasSibiryakovPRL} (its Hamiltonian
formulation discussed in papers \cite{KlusonJHEP, KlusonPRD})) was created to
cure some problems of Ho\v{r}ava-type models, but the proposed cure leads to
side effects which are far more severe than the original illness (if not terminal).

\end{document}